\documentclass[%
 reprint,
superscriptaddress,
nofootinbib,
 amsmath,amssymb,
 aps,
]{revtex4-2}
\usepackage{graphics,epsfig,psfrag}
\usepackage{color}
\usepackage{graphicx}
\usepackage{dcolumn}
\usepackage{bm}

\begin{document}

\title{Analytical study of higher-order ring images\\
of accretion disk around black hole}

\author{Gennady S. Bisnovatyi-Kogan}
\email{gkogan@iki.rssi.ru}
\affiliation{Space Research Institute of Russian Academy of Sciences, Profsoyuznaya 84/32, Moscow 117997, Russia}
\affiliation{National Research Nuclear University MEPhI (Moscow Engineering Physics Institute), Kashirskoe Shosse 31, Moscow 115409, Russia}
\affiliation{Moscow Institute of Physics and Technology, 9 Institutskiy per., Dolgoprudny, Moscow Region, 141701, Russia}%

\author{Oleg Yu. Tsupko}
\email{tsupko@iki.rssi.ru; tsupkooleg@gmail.com}
\affiliation{Space Research Institute of Russian Academy of Sciences, Profsoyuznaya 84/32, Moscow 117997, Russia}

\date{\today}

\begin{abstract}
Gravitational lensing of a light source by a black hole leads to appearance of higher-order images produced by photons that loop around the black hole before reaching the observer. Higher-order images were widely investigated numerically and analytically, in particular using so-called strong deflection limit of gravitational deflection. After recent observations of the black hole image, attention has been drawn to higher-order rings, which are lensed images of the accreting matter of the black hole environment and can appear near the boundary of the black hole shadow. In this article, we use strong deflection limit technique to investigate  higher-order ring images of luminous accretion disc around a Schwarzschild black hole. We consider thin disk given by the inner and outer radii and an observer located far from the black hole on the axis of symmetry. For this configuration, it becomes possible to find the angular radii, thicknesses, and solid angles of higher-order rings in the form of compact analytical expressions. We show that the size of the rings is mainly determined by the position of the inner boundary of the accretion disk, which makes it possible to use them to distinguish between different accretion models. Possible generalizations of our model to non-symmetric images can help to make the estimation of black hole angular momentum. We also present the analytical estimation of fluxes from higher-order images. Our method makes it easy to investigate $n=2$ and $n=3$ higher-order rings, the possible observation of which in future projects is currently being discussed. 
\end{abstract}

\maketitle

\section{Introduction}

One of the most famous effects of General Relativity is the deflection of light by massive bodies. Such a gravitating body is often called a gravitational lens. Light from a source, deflected by a gravitational lens, can reach the observer in several ways, which leads to an amazing phenomenon -- the appearance of multiple images of the same source.

If the gravitating body is a black hole, then the light rays can move in a close vicinity of the gravitational radius. In this case, the angles of deflection of the light rays can be very large. In particular, the photons can make one or more revolutions around the black hole before reaching the observer. Such photons give rise to so-called higher-order images (also known as relativistic images) of distant source.

Studies of higher-order images produced by photons that orbit a black hole have a long history. Usually, images of a source that is far from the black hole were considered. Such images were investigated in articles by Darwin \cite{Darwin-1959} and Ohanian \cite{Ohanian-1987}. They are also discussed in Misner, Thorne and Wheeler's textbook \cite{MTW-1973}.

The active study of higher-order images began about two decades ago. Virbhadra and Ellis calculated numerically the properties of higher-order images in the case of lensing by Schwarzschild black hole \cite{Virbhadra-2000}, see also \cite{Virbhadra-2001, Virbhadra-2009}. To denote higher-order images, they introduced the term 'relativistic images' which was often used later on. An important contribution was made in a series of works by Bozza with co-authors \cite{Bozza-2001, Bozza-2002, Bozza-2003, Bozza-2005, Bozza-Mancini-ApJ-2004, Bozza-Mancini-ApJ-2005, Bozza-2006, Bozza-Sereno-2006, Bozza-Scarpetta-2007} who calculated the size and the magnification of relativistic images analytically for spherically symmetric and Kerr black holes. They used so called strong deflection limit of gravitational deflection: analytical logarithmic expression for the deflection angle valid for light rays that have made one or more loops around the black hole. Strong deflection limit was used also in series of papers of Eiroa with co-authors \cite{Eiroa-2002, Eiroa-2004, Eiroa-2005, Eiroa-2006}. Different types of lens equations were proposed for studying higher-order images in works of Frittelli et al \cite{Frittelli-2000}, Virbhadra and Ellis \cite{Virbhadra-2000}, Bozza et al \cite{Bozza-2001, Bozza-2002, Bozza-Sereno-2006, Bozza-2008}, Perlick \cite{Perlick-2004a, Perlick-2004b}, Aazami et al \cite{Aazami-2011a, Aazami-2011b}.

Since then, there have been many works that have investigated higher-order images, both numerically and analytically, for some examples, see \cite{Amore-2007, Iyer-Petters-2007, Gyulchev-2007, Chen-Jing-2009, Ghosh-2010, Bin-Nun-2011, Ding-Kang-2011, Wei-2012, Tsupko-BK-2013, Sadeghi-2014, Alhamzawi-2016, Tsukamoto-2016, Chakraborty-2017, Barlow-2017, Bozza-2017, Uniyal-2018, Bergliaffa-2020, Tsukamoto-2021a, Tsukamoto-2021b, Islam-2021, Furtado-2021, Aratore-Bozza-2021}. Analytical logarithmic limit of gravitational deflection is commonly referred to as 'strong deflection limit' (sometimes 'strong field limit'). More generally, for lensing research beyond the weak deflection approximation, the terms 'strong deflection gravitational lensing' and 'strong gravitational lensing by black hole' are used. In the latter case, these studies should not be confused with strong lensing in conventional observational gravitational lensing: when multiple images are formed (i.e. the lensing effect is 'strong'), but the deflection angles are still small. For review of gravitational lensing beyond the weak deflection approximation, see, e.g., \cite{Perlick-2004a, Bozza-2010, BK-Tsupko-Universe-2017}.

An interesting special case arises when a distant source, a black hole and an observer are perfectly aligned. In this case, an infinite sequence of concentric higher-order ring images around black hole is formed (see, e.g., \cite{Virbhadra-2000, Bozza-2001, Bozza-2002, BK-Tsupko-2008}), sometimes referred as 'Einstein relativistic rings' \cite{Virbhadra-2000}. Study of such rings for the compact distant source of given angular size can be found in \cite{BK-Tsupko-2008}. Term 'Einstein-Chwolson ring' is also used, see original paper \cite{Chwolson-1924}.

After recent observation of black hole shadow in M87 \cite{Falcke-2000, EHT-1, EHT-2, EHT-3, EHT-4, EHT-5, EHT-6, Bronzwaer-Falcke-2021}, the attention has been drawn to investigation of higher-order rings which can arise around the black hole shadow and can be probably observed in the substructure of black hole image \cite{Gralla-2019, Johnson-2020, Gralla-Lupsasca-2020, Gralla-Lupsasca-Marrone-2020, Broderick-2021, Pesce-2021, Wielgus-2021}. These rings are lensed images of luminous accreting matter of the black hole environment. A detailed numerical and analytical discussion had been presented. Due to the assumed angle of observation in the galaxy M87, of particular interest is the case of a polar observer who sees the accretion disk face-on. Observation of high-order rings in future projects is now actively discussed, e.g. \cite{Johnson-2020, Gralla-Lupsasca-Marrone-2020, Broderick-2021, Pesce-2021}.

\begin{figure*}
\begin{center}
\includegraphics[width=0.85\textwidth]{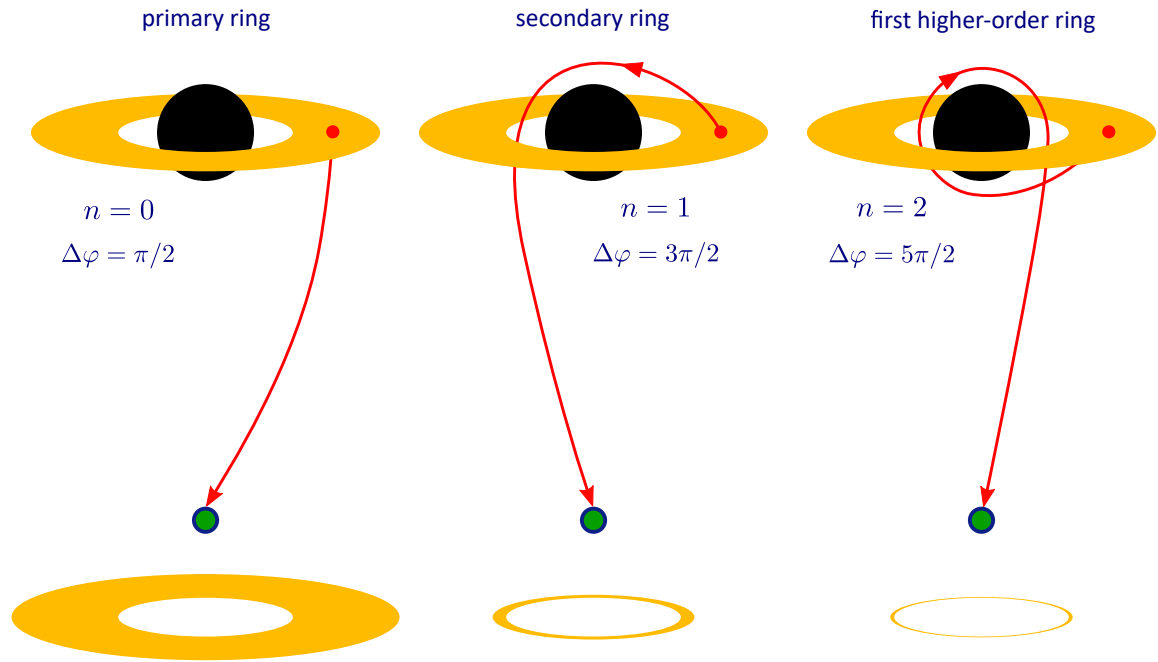}
\end{center}
\caption{Images of thin accretion disk in case of lensing by spherically symmetric black hole and an observer located on the axis of symmetry. For every image we write the number $n$ of half-orbits \cite{Johnson-2020, Broderick-2021, Pesce-2021, Wielgus-2021} and the change of angular coordinate $\Delta \varphi$ of the light ray, see Section \ref{sec: nomenclature} for more details. The objects on the figure are not in scale. An idea of the real dimensions of the rings in relation to each other can be obtained from Fig.\ref{fig:pr-sec-6-15} in Section \ref{sec: example}.}
\label{fig:rings}
\end{figure*}

In this paper, we apply earlier developed formalism of strong deflection gravitational lensing in novel studies of higher-order ring images of luminous accretion matter around the Schwarzschild black hole. In particular, we use strong deflection limit formulas for arbitrary source position \cite{Bozza-Scarpetta-2007, Bozza-2010, Aratore-Bozza-2021}.

Our goal is to derive a fully analytical solution, for a simplified case. This can help to reveal new features of the problem that may not be visible in a numerical calculation or in more detailed analytical studies where more complicated account of the parameters is used.

Here we consider a thin luminous accretion disk with inner and outer radii around the Schwarzschild black hole, and the observer who is located on the axis of symmetry perpendicular to the equatorial plane. Observer will see the primary (direct) image of disk, a secondary image in the form of a thin ring (image of the back of the disk), and a sequence of exponentially weak higher-order rings (Fig.\ref{fig:rings}). For this configuration, we present fully analytical calculation of higher-order rings properties. Radii, thicknesses and solid angles of higher-order rings are found in the form of compact analytical expressions. The resulting solution is analyzed in detail. We also present the analytical estimation of fluxes of the rings.

The paper is organized as follows. In the next Section, we briefly introduce the notion of strong deflection limit. Then, in Section \ref{sec: nomenclature} we describe and classify lensed images of the accretion disk, with relation to nomenclature of images of relativistic rings already used in the literature. In Section \ref{sec: derivation} we derive the analytical expressions which describe the higher-order rings: angular thicknesses of rings and solid angles occupied by ring images. In Section \ref{sec: properties} we investigate the properties of rings. In particular, we explore the dependence of solution on position of inner boundary of accretion disk. In Section \ref{sec: flux} the analytical estimation of fluxes and magnifications is discussed. In Section \ref{sec: example} we provide an example of calculation for specific values of inner and outer radii, and also draw the picture of first three rings together, see Fig.\ref{fig:pr-sec-6-15}. Section \ref{sec: conclusions} is Conclusions.

\section{Strong deflection limit: brief introduction}

Here we briefly remind what the strong deflection limit is and how it is written analytically in the simplest situation.

We write the Schwarzschild metric as
\begin{equation}  \label{schw-metric}
ds^2 = - \left(1 - \frac{2m}{r} \right) \, c^2 dt^2 + \frac{dr^2}{1-2m/r} \, +
\end{equation}
\[
+ \, r^2
\left( d \vartheta^2 + \sin^2 \vartheta \, d \varphi^2 \right),   \;   m=\frac{GM}{c^2}   \,   ,
\]
where $m$ is a mass parameter of dimension of length, and $M$ is the black hole mass. In these notations, we have:
\begin{itemize}
    \item horizon at radial coordinate $2m$;
    
    \item photon sphere filled by unstable circular orbits at radial coordinate $r_{\mathrm{ph}}=3m$;
    
    \item critical value $b_\mathrm{cr} = 3\sqrt{3}m$ of the impact parameter separating the captured and flyby orbits of light rays coming from infinity; this value also determines the angular radius of black hole shadow for the observer at large distances, $\alpha_{\mathrm{sh}} = 3\sqrt{3}m/D_d$, where $D_d$ is the distance between the black hole and the observer;
    
    \item innermost stable circular orbit for massive particles at $r_\mathrm{ISCO} = 6m$.
\end{itemize}

Let us consider a light ray that moves from infinity to a black hole, reaches the minimum value of the radial coordinate along the trajectory $R$ (usually referred as the distance of the closest approach) and then flies away to infinity. Change of azimuthal angle for such light ray equals to (e.g., \cite{Hartle-2003, Congdon-Keeton-2018}):
\begin{equation}
\Delta \varphi = 2 \int \limits_R^\infty \frac{1}{r^2}
\left[ \frac{1}{b^2} - \frac{1}{r^2} \left( 1- \frac{2m}{r}	\right) \right]^{-1/2} dr .
\end{equation}
The impact parameter $b$, corresponding to the distance of closest approach $R$, is written as
\begin{equation}
\label{b-and-R} b^2 = \frac{R^3}{R-2m} \, .
\end{equation}

The change in the angular coordinate in the case of a straight light ray in the absence of deflection equals to $\pi$. Therefore the deflection angle, which is the angle between the asymptotic incoming and outgoing directions of the ray, is computed as:
\begin{equation}
\label{vacuum-exact}
\hat{\alpha} = 2 \int \limits_R^\infty \frac{1}{r^2}
\left[ \frac{1}{b^2} - \frac{1}{r^2} \left( 1- \frac{2m}{r}	\right) \right]^{-1/2} dr \,  - \,  \pi \, .
\end{equation}

If the impact parameter is large, $b \gg m$, then the angle of deflection is small, $\hat{\alpha} \ll 1$, and can be computed by the formula of the Einstein angle:
\begin{equation} \label{einst-b}
\hat{\alpha} \, = \, \frac{4m}{R} \, , \quad \mbox{or} \quad \hat{\alpha} \, = \, \frac{4m}{b}  \,   .
\end{equation}

There is also another limiting case when the deflection angle can be written analytically: the strong deflection limit (also sometimes referred as strong field limit). Physically, the strong deflection limit corresponds to a situation when a photon approaches a black hole, makes one or several loops around it, and then flies away. In this case, the deflection angle is large: $\hat{\alpha} \gg 1$.

The deflection angle of photon in strong deflection limit is written as \cite{Darwin-1959, Bozza-2001}:
\begin{equation}
\label{vacuum-alpha-R}
\hat{\alpha} =  - 2 \ln \frac{R-3m}{36(2-\sqrt{3}) m} - \pi \, ,
\end{equation}
or, as a function of the impact parameter $b$ \cite{Bozza-2001, Bozza-2002}:
\begin{eqnarray}
\label{vacuum-alpha-b}  \hat{\alpha} &=& - \ln \left( \frac{b}{b_{cr}} - 1 \right) + \ln[216(7-4\sqrt{3})] - \pi \, = \\
&=& - \ln \frac{b-3\sqrt{3}m}{648\sqrt{3}(7-4\sqrt{3})m} - \pi = \nonumber\\
&=& - \ln \frac{(b-3\sqrt{3}m)(7\sqrt{3}+12)}{1944m} - \pi \, .\nonumber
\end{eqnarray}

Strong deflection limit for deflection angle of massive particles in Schwarzschild metric has been first derived by Tsupko \cite{Tsupko-2014}, see also \cite{Liu-2016, Crisnejo-Gallo-Jusufi-2019}.

Note that the formula (\ref{vacuum-alpha-b}) cannot be directly applied for our purposes, because we are interested a situation where the source (emitting accretion disk, and especially its inner parts) is in the immediate vicinity of a black hole. We will use a more general formula suitable for an arbitrary position of the source, see Section \ref{sec: derivation}.

\section{Lensing of accretion disk by a black hole}
\label{sec: nomenclature}

Luminet \cite{Luminet-1979} presented a visual appearance of a thin emitting accretion disk around Schwarzschild black hole, for an observer located slightly above the equatorial plane, see Fig.11 there. In that picture, there are, as he called them, 'direct (or primary)' and 'ghost (or secundary)' images of accretion disc found numerically. Higher-order images were discussed in the paper \cite{Luminet-1979}  but were not presented at the picture.

In Luminet's picture, both primary and secondary images of disk are strongly deformed. As a result, we see only top part of secondary image which looks like thin arc around central dark spot (shadow). The lower part of secondary image is hidden by primary image. For good illustration of formation of this image see Fig.3 in \cite{Luminet-2019}.

Due to the assumed angle of observation in M87, special attention has been now drawn to the case where the observer is assumed to be close to the axis of rotation \cite{Gralla-2019, Broderick-2021}. For the observer located on the axis symmetry, secondary image will have a form of non-deformed thin ring. There will be also infinite series of weak higher-order rings which are concentrated near the boundary of black hole shadow. The entire system of all ring images (including secondary image or not) is now often referred as just 'photon ring'. Sometimes the notion 'photon ring' is also used for each individual ring.

In this paper, we will consider thin accretion disk around Schwarzschild black hole and the observer located on the axis of symmetry (polar observer). For naming the lensed images of accretion disk, we will follow the standard terminology of gravitational lensing, which is also consistent with the works of Luminet \cite{Luminet-1979} and Broderick et al \cite{Broderick-2021}. For numbering of the images, we will follow the commonly used notation of \cite{Johnson-2020, Broderick-2021, Pesce-2021, Wielgus-2021}. The number $n$ approximately corresponds to the number of half-turns that the ray makes, moving from the source to the observer.

We have the following images  of accretion disk (Fig.\ref{fig:rings}):

\begin{itemize}

\item Primary image ($n=0$). It is direct image of accretion disc slightly increased by the gravitational bending of light rays. 

\item Secondary image ($n=1$) in the form of thin ring. It is image of the back of the disk formed by photons that have made about half a turn on their way to the observer \cite{Gralla-2019, Johnson-2020, Broderick-2021}.

\item Higher-order images ($n\ge2$). These are exponentially weak rings formed by photons that have made one full turn or more around the black hole. Rings are concentrated very close to the boundary of black hole shadow.

\end{itemize}


The importance of the secondary image over the following higher-order rings has been discussed in Gralla et al. \cite{Gralla-2019} (who call it the 'lensing ring') and Johnson et al. \cite{Johnson-2020} (who call it the 'leading subring'). As stated, secondary ring gives the leading contribution to thin ring-like structure in black hole image, while higher-order images are exponentially weaker. For further discussion, it is important to emphasize that the secondary image is not exponentially weak and is poorly described by the strong deflection limit. Exponential character of higher-order images in the case of accretion disk lensing has been discussed in \cite{Gralla-2019, Johnson-2020}. Comprehensive analytical approach for higher-order images has been developed in \cite{Gralla-Lupsasca-2020}. Higher-order rings in spherically symmetric metrics have been discussed by Wielgus \cite{Wielgus-2021} in a simple model of infinitely thin ring of given radius. Particular attention is paid to discussing the possibility of observing the next rings after the secondary image \cite{Johnson-2020, Gralla-Lupsasca-Marrone-2020, Broderick-2021, Pesce-2021}.

We emphasize that in this work, we are examining higher-order rings that are very close to the boundary of shadow, but not the shadow itself. The size and the shape of the shadow boundary can be described analytically, see e.g. \cite{Synge-1966, Zeld-Novikov-1965, Bardeen-1973, Chandra-1983, Dymnikova-1986, Stuchlik-1999, Takahashi-2004, Zakharov-Paolis-2005-New-Astronomy, Johannsen-2010, Johannsen-2013, Gren-Perlick-2014, Gren-Perlick-2015, Perlick-Tsupko-BK-2015, Cunha-2015, Abdujabbarov-2015, Shipley-Dolan-2016, Tsupko-2017, Perlick-Tsupko-2017, Yan-2018, Yunes-2018, Mars-2018, Cunha-2018, Perlick-Tsupko-BK-2018, BK-Tsupko-2018, Wei-2019-Rapid, BK-2019, Abdikamalov-2019, Psaltis-2019-review}. For review, see Cunha and Herdeiro \cite{Cunha-Herdeiro-2018} and Perlick and Tsupko \cite{Perlick-Tsupko-2022}. To obtain a realistic visual appearance of shadow together with accreting environment, it is necessary to carry out numerical simulations including ray tracing, e.g. \cite{Falcke-2000, Broderick-Loeb-2005, Moscibrodzka-2009, Dexter-2009, Broderick-Loeb-2009, Broderick-Fish-2009, Broderick-Johannsen-2014, James-2015, Mizuno-2018, Narayan-2019}.
For studies of 2020--2021 related to black hole shadow see, e.g., \cite{Tsupko-Fan-BK-2020, Vagnozzi-2020, Neves-2020, Farah-2020, Li-Guo-2020, Chang-Zhu-2020, Wielgus-2020, Alexeyev-2020, Tsupko-BK-2020-IJMPD, Kumar-Ghosh-2020, Cunha-2020, Lima-2020, Dokuchaev-2020, Chael-2021, Bronnikov-2021, Pantig-2021, Anacleto-2021, Devi-2021, Guerrero-2021, Lima-2021, Frion-2021, Khodadi-2021, Tsupko-2021, Cardoso-2021, Psaltis-EHT-2020, Andrianov-2021, Bronzwaer-2021, EHT-2021, Wang-2021, Eichhorn-2021, Ozel-2021, Roy-2021}.

\section{Analytical calculation of higher-order ring images of accretion disk}
\label{sec: derivation}

In this Section we calculate analytically  the properties of higher-order ring images of thin accretion disk. The disk is given by the inner and outer radii. The observer is located on the axis of symmetry at a large distance from the black hole (much larger than its gravitational radius), see Fig.\ref{fig:rings}.
We remind that number $n$ of half-orbits numerate all images starting from the primary ($n=0$) and secondary ($n=1$), whereas the results of this Section are applicable only for higher-order images ($n\ge2$).

For calculation of the higher-order images, we use the strong deflection limit technique. We would like to note two features of the problem under consideration:

(i) Most of the previous work dealing with the strong deflection approximation has considered sources that are far from the black hole. In our paper, the radiation sources are located in immediate vicinity of black hole where its gravitational field cannot be neglected. Strong deflection limit of black hole gravitational lensing was generalized to the case of arbitrary source positions by Bozza and Scarpetta \cite{Bozza-Scarpetta-2007}, see also \cite{Bozza-2010, Aratore-Bozza-2021}. Instead of the deflection angle $\hat{\alpha}$, a change in the angular coordinate of the ray (or azimuthal shift) $\Delta \varphi$ is used.

(ii) Previous work was mainly focused on the compact distant source. In our article, considering the accretion disk, we are dealing with an extended distribution of light sources. Therefore, to compute the parameters of images, it is necessary to integrate over the image surface. However, thanks to our choice of a simplified symmetric configuration, we can find the solid angle of the higher-order ring by calculating its outer and inner angular radii.\\

\begin{figure}
\begin{center}\vspace{20pt}
\includegraphics[width=0.45\textwidth]{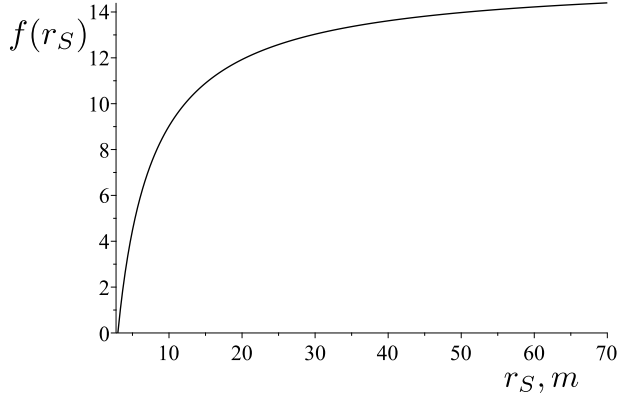}
\end{center}
\caption{Graph of the function $f(r_S)$ given by eq.(\ref{f-rS}).}
\label{fig:graph0}
\end{figure}

Before handling with a luminous disk, let us first consider a point source with an arbitrary position outside the photon sphere. Radial coordinate of the source is $r_S$. We are interested in rays that move from the source, approach the black hole, make one or several revolutions around the black hole while reaching the distance of the closest approach $R$, and then fly away to infinity. These rays are responsible for formation of higher-order images of source for an observer at a great distance. The change of the angular coordinate $\Delta \varphi$ of such ray in strong deflection limit is written as \cite{Bozza-Scarpetta-2007, Bozza-2010, Aratore-Bozza-2021}:
\begin{equation} \label{bozza-2007-03}
\Delta \varphi = - \ln \epsilon
+ \ln f(r_S) \, ,
\end{equation}
where
\begin{equation} \label{bozza-2007-04}
\epsilon = \frac{b - b_{cr}}{b_{cr}} \ll 1 \, , \quad b_{cr} = 3\sqrt{3}m \, ,
\end{equation}
\begin{equation} \label{f-rS}
f(r_S) =  \frac{6^5 \left(1 - \frac{3m}{r_S} \right) }{
\left(3+\sqrt{3} \right)^2 
} \left( 3+ \sqrt{3 + \frac{18m}{r_S}} \right)^{-2} \, .
\end{equation}
Graph of the function $f(r_S)$ is plotted on Fig.\ref{fig:graph0}. 
In the limit $r_S \gg m$,
\begin{equation} \label{f0}
f(r_S) \to f_0 \equiv  \frac{7776}{(3+\sqrt{3})^4} = 216 \, (7 - 4\sqrt{3}) \simeq 15.5  \, ,
\end{equation}
the deflection angle can be introduced in a usual way $\hat{\alpha} = \Delta \varphi - \pi$, obtaining the formula (\ref{vacuum-alpha-b}) for the deflection angle $\hat{\alpha}$. \\

We start from eq.(\ref{bozza-2007-03}) and find
\begin{equation} \label{b-general}
b = b_{cr} \left[ 1 + f(r_S) \,  e^{-\Delta \varphi}   \right] \, .
\end{equation}

Since the observer is in an asymptotically flat region, the angular position of image for observer and the impact parameter are related by a simple relationship \cite{Bozza-Scarpetta-2007}: $\theta=b/D_d$. We obtain than
\begin{equation} \label{theta-general}
\theta = \frac{b_{cr}}{D_d} \left[ 1 + f(r_S) \,  e^{-\Delta \varphi}   \right] \, .
\end{equation}
This formula describes the observed angular position $\theta$ of higher-order image formed by the light ray that has experienced the change in the angular coordinate $\Delta \varphi$ when moving from the point source with coordinate $r_S$ to the observer located at distance $D_d \gg m$ from the black hole.

To apply eq.(\ref{theta-general}) to consideration of thin luminous disk viewed face-on, we have to know the corresponding values of $\Delta \varphi$ for every image. From the Fig.\ref{fig:rings} we can write the general rule as
\begin{equation} \label{Delta-phi}
\Delta \varphi =  \pi \left( n  + \frac{1}{2} \right) \, , \quad n \ge 0.
\end{equation}
Note that the formula (\ref{theta-general}) can be applied only for higher-order images ($n\ge2$).

\begin{figure}
\begin{center}
\includegraphics[width=0.45\textwidth]{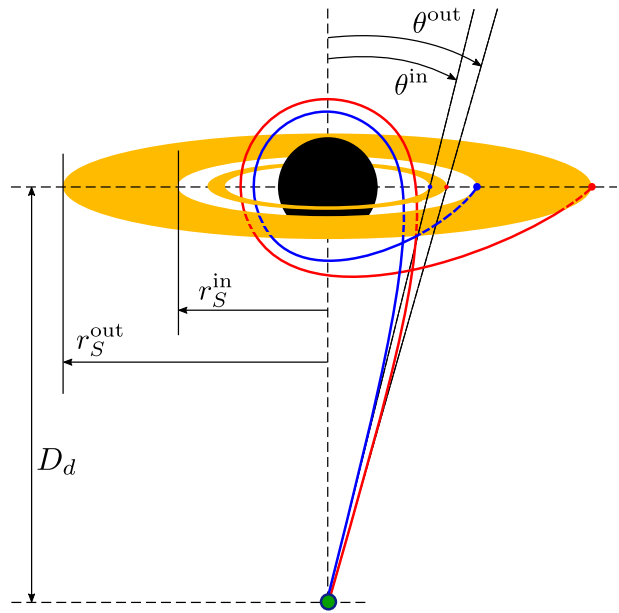}
\end{center}
\caption{Formation of $n=2$ higher-order ring in the case of gravitational lensing of the accretion disk by Schwarzschild black hole. This is the next ring after the secondary image ($n=1$), which is also ring-shaped. The observer is located on the axis of symmetry at a great distance $D_d \gg m$ from the black hole. The accretion disk is geometrically thin and is defined by two radii: inner $r_S^\mathrm{in}$ and outer $r_S^\mathrm{out}$. Two rays are shown in different colors, coming from the inner and outer edges of the disk. Such rays make approximately one revolution around the black hole and form an annular image with the inner angular radius $\theta^\mathrm{in}$ and the outer angular radius $\theta^\mathrm{out}$. The solid angle occupied by this ring can be found fully analytically with our eq.(\ref{Omega-n}) by substitution $n=2$. The objects on the figure are not in scale. In particular, $n=2$ ring is much thinner. For an illustration of the actual dimensions, see Fig.\ref{fig:pr-sec-6-15}.}
\label{fig:n2-ring}
\end{figure}

We set the size of accretion disk with an inner radius $r_S^{\mathrm{in}}$ and an outer radius $r_S^{\mathrm{out}}$ (Fig.\ref{fig:n2-ring}). Corresponding inner and outer angular radii of higher-order ring image with given $n$ are then:
\begin{equation} \label{theta-in}
\theta^{\mathrm{in}}_n = \frac{3\sqrt{3}m}{D_d} \left[ 1 + f(r_S^{\mathrm{in}}) \,  e^{- \pi \left( n+ \frac{1}{2} \right)}  \right] \, ,
\end{equation}
\begin{equation} \label{theta-out}
\theta^{\mathrm{out}}_n = \frac{3\sqrt{3}m}{D_d} \left[ 1 + f(r_S^{\mathrm{out}}) \,  e^{- \pi \left( n+ \frac{1}{2} \right)}   \right] \, .
\end{equation}
The angular thickness $\Delta \theta_n \equiv \theta^{\mathrm{out}}_n - \theta^{\mathrm{in}}_n$ of the $n$-ring equals to:
\begin{equation} \label{theta-n}
\Delta \theta_n  =  \frac{3\sqrt{3}m}{D_d}  \left[  f(r_S^{\mathrm{out}}) -  f(r_S^{\mathrm{in}})    \right]
e^{- \pi \left( n+ \frac{1}{2} \right)}  \, .
\end{equation}

In the same manner as we did for calculation of higher-order rings of distant compact source \cite{BK-Tsupko-2008}, we write the angular size of higher-order ring ($n \ge 2$) as the difference of solid angles occupied by cone with angular size $\theta^{\mathrm{out}}_n$ and cone with angular size $\theta^{\mathrm{in}}_n$:
\begin{equation}
\Delta \Omega_n = \pi  \left[ (\theta^{\mathrm{out}}_n)^2 - (\theta^{\mathrm{in}}_n)^2  \right] =
\end{equation}
\[
= \pi \left( \theta^{\mathrm{out}}_n - \theta^{\mathrm{in}}_n  \right)
\left( \theta^{\mathrm{out}}_n + \theta^{\mathrm{in}}_n  \right) \, .
\]

From (\ref{bozza-2007-04}) and (\ref{b-general}), we see that $f(r_\mathrm{S}) \exp(-\Delta \varphi) \ll 1$, therefore
\begin{equation}
\theta^{\mathrm{out}}_n + \theta^{\mathrm{in}}_n \simeq \frac{2 \cdot 3\sqrt{3}m}{D_d} \, .
\end{equation}
Finally we obtain the solid angle of higher-order images as
\begin{equation} \label{Omega-n}
\Delta \Omega_n = 2 \pi \, \frac{27m^2}{D_d^2}   \left[  f(r_S^{\mathrm{out}}) -  f(r_S^{\mathrm{in}})    \right]    e^{- \pi \left( n+ \frac{1}{2} \right)} \, , \; n \ge 2.
\end{equation}
We emphasize that the method works well only if condition $\epsilon \ll 1$ holds, see eq.(\ref{bozza-2007-04}). It means that it should be $f(r_\mathrm{S}) \exp(-\pi \left( n+ 1/2 \right)) \ll 1$. It is easy to check that this condition is satisfied already for $n = 2$.

\section{Properties of solution found}
\label{sec: properties}

In previous Section we have found the compact expressions (\ref{theta-in}), (\ref{theta-out}), (\ref{theta-n}), (\ref{Omega-n}) which allows one to calculate easily the properties of higher-order rings in the case of lensing of accretion disk by black hole. Since all relationships are written in an analytical form, this allows us to analyze their properties and explore various dependencies. We will do this in this Section.

(i) Very important property is that the thickness of the rings decreases exponentially:
\begin{equation}
\Delta \theta_{n+1} = e^{-\pi} \Delta \theta_n    \, , \quad n\ge2 \, . 
\end{equation}
The angular radius of all rings remains approximately the same (slightly larger than the size of the shadow $3\sqrt{3}m/D_d$). The solid angle occupied by the ring also decreases exponentially, see eq.(\ref{Omega-n}):
\begin{equation}
\Delta \Omega_{n+1} = e^{-\pi}  \Delta \Omega_n \, , \quad n\ge2 \, .
\end{equation}
This connection between two subsequent higher-order rings has already been highlighted in the literature \cite{Gralla-2019, Gralla-Lupsasca-2020, Johnson-2020}. However, in our work we not only reproduce this property, but also find analytically the size of each of the rings.

(ii) The solid angle of the sum of all higher-order images is calculated as
\begin{equation}
\Omega_{\mathrm{n\ge2}} = \sum \limits_{n=2}^\infty \Delta \Omega_n =
\end{equation}
\[
= 2 \pi \, \frac{27m^2}{D_d^2}   \left[  f(r_S^{\mathrm{out}}) -  f(r_S^{\mathrm{in}})    \right] \frac{e^{-5\pi/2}}{1 - e^{-\pi}}  \, .
\]

(iii) Consider the dependence of solution on position of inner boundary of accretion disk.

Formulas for angular thickness (\ref{theta-n}) and for solid angle (\ref{Omega-n}) contain dependence on the inner $r_S^{\mathrm{in}}$ and outer $r_S^{\mathrm{out}}$ radii of the disk only as a combination of functions $f(r_S)$ inside square brackets.

Let us now assume that $r_S^{\mathrm{out}} \gg m$, whereas $r_S^{\mathrm{in}}$ is of the order of several $m$. Then the solid angle of the unlensed disk $\Delta \Omega_S$ is determined by the outer radius mainly:
\begin{equation}
\Delta \Omega_S = \frac{\pi}{D_d^2}  \left[ (r^{\mathrm{out}}_S)^2 - (r^{\mathrm{in}}_n)^2  \right] \simeq  \frac{\pi}{D_d^2}  (r^{\mathrm{out}}_S)^2  \, .
\end{equation}

\begin{figure}
\begin{center}\vspace{20pt}
\includegraphics[width=0.45\textwidth]{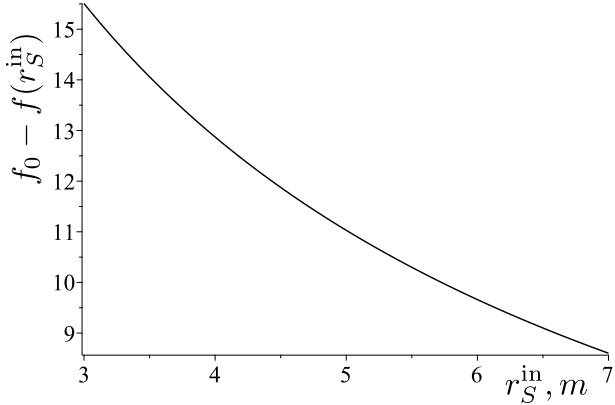}
\end{center}
\caption{Expression in square brackets $[f_0 -  f(r_S^{\mathrm{in}})]$ in (\ref{theta-n-new}) and (\ref{Omega-n-new}) as the function of inner border of the disc $r_S^{\mathrm{in}}$.}
\label{fig:graph1}
\end{figure}

In contrast, a size of higher-order images depends mainly on the inner radius $r_S^{\mathrm{in}}$ of the accretion disk. Indeed, for the function $f(r_S^{\mathrm{out}})$ used for higher-order images we can write (see  eq.(\ref{f0}))
\begin{equation}
f(r_S^{\mathrm{out}}) \simeq f_0 \equiv  \frac{7776}{(3+\sqrt{3})^4} = 216 \, (7 - 4\sqrt{3})   \, ,
\end{equation}
Then the expression in square brackets in (\ref{theta-n}) and (\ref{Omega-n}) can be simplified, and we write the angular thickness and the solid angle of image with given a number $n$ as the function of the inner radius of accretion disk only:
\begin{equation} \label{theta-n-new}
\Delta \theta_n (r_S^{\mathrm{in}})  \simeq  \frac{3\sqrt{3}m}{D_d} \left[  f_0 -  f(r_S^{\mathrm{in}})    \right]
e^{- \pi \left( n+ \frac{1}{2} \right)} \, ,
\end{equation}
\begin{equation} \label{Omega-n-new}
\Delta \Omega_n (r_S^{\mathrm{in}})   \simeq   2 \pi \, \frac{27m^2}{D_d^2}   \left[  f_0 -  f(r_S^{\mathrm{in}})    \right]    e^{- \pi \left( n+ \frac{1}{2} \right)} \, , \; n \ge 2.
\end{equation}
In Fig.\ref{fig:graph1} we plot the expression in the square brackets as the function of $r_S^{\mathrm{in}}$. We see that the dependence of the size of the rings on the position of the inner boundary of the disk is quite strong.

\begin{figure}
\begin{center}
\includegraphics[width=0.46\textwidth]{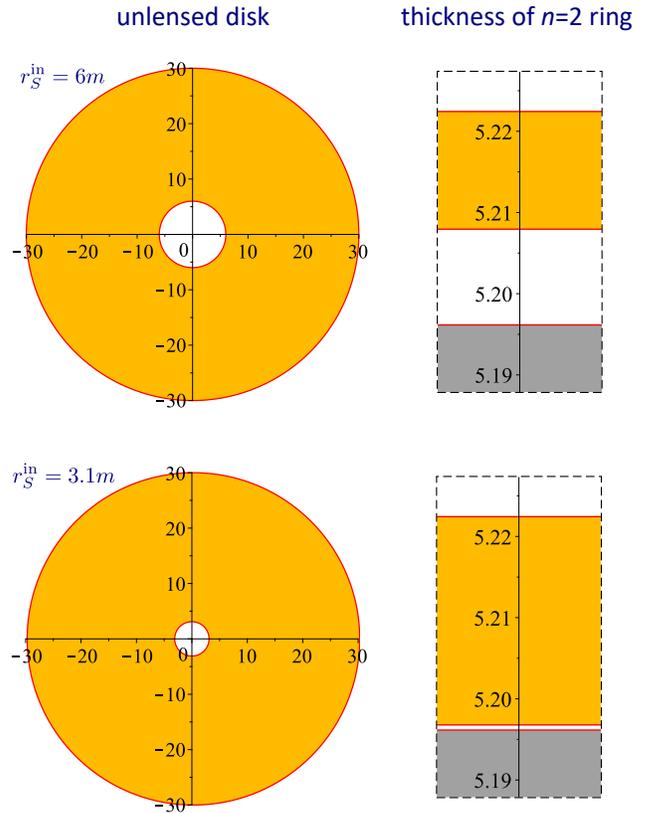}
\end{center}
\caption{Strong dependence of the observed angular thickness $\Delta \theta_n$ of higher-order ring on position of the inner boundary $r_S^{\mathrm{in}}$ of accretion disk. Top: In the left column we show the \textit{unlensed} size of accretion disk with $r_S^{\mathrm{in}} = 6m$ and $r_S^{\mathrm{out}} = 30m$. Images $n=0$ and $n=1$ are not presented here; compare with Fig.\ref{fig:pr-sec-6-15}. In the right column we show the thickness of $n=2$ higher-order ring. Gray area is the part of black hole shadow. Inner and outer angular radii of the higher-order image are calculated with Eqs. (\ref{theta-in}) and (\ref{theta-out}) correspondingly. Bottom: The same picture but with another inner boundary of the disk, $r_S^{\mathrm{in}} = 3.1m$. The solid angle of the non-lensed disk changes very little in comparison with the previous case. At the same time, the thickness of the higher-order ring grows quite significantly. }
\label{fig:inner}
\end{figure}

It could be noticed in Fig.\ref{fig:inner}. If we change the position of the inner boundary of disk $r_S^{\mathrm{in}}$ from $6m$ (position of ISCO) to $3.1m$ (slightly bigger than the position of the photon sphere), then the angular size of the disk itself increases little, since its outer boundary is large $r_S^{\mathrm{out}} \gg m$. At the same time, the thickness and the solid angle of each higher-order ring will increase significantly. This allows us to relate the assumed accretion disk model to the higher-order ring size in a simple way.

(iv) The importance of the inner layers of the disc in comparison with the outer ones can also be shown by the following reasoning. Let us consider a source as a thin ring with radius $r_S$ and with thickness $\Delta r_S \ll m$. We write:
\begin{equation}
f(r_S^{\mathrm{out}}) -  f(r_S^{\mathrm{in}})   \simeq
f_1(r_S) \, \frac{\Delta r_S}{m} \, ,
\end{equation}
where
\begin{equation}
f_1(r_S) \equiv
\left. m \, \frac{df(r)}{dr} \right|_{r=r_S}  \, .
\end{equation}
Differentiating eq.(\ref{f-rS}), we find:
\begin{equation} \label{f1-rS}
f_1(r_S) =  \frac{23328 \sqrt{3} \, m^2}
{
r^2  \left(3+\sqrt{3} \right)^2  \sqrt{1 + \frac{6m}{r}} 
}
\left( 3+ \sqrt{3 + \frac{18m}{r_S}} \right)^{-2} \, .
\end{equation}

\begin{figure}
\begin{center}\vspace{20pt}
\includegraphics[width=0.45\textwidth]{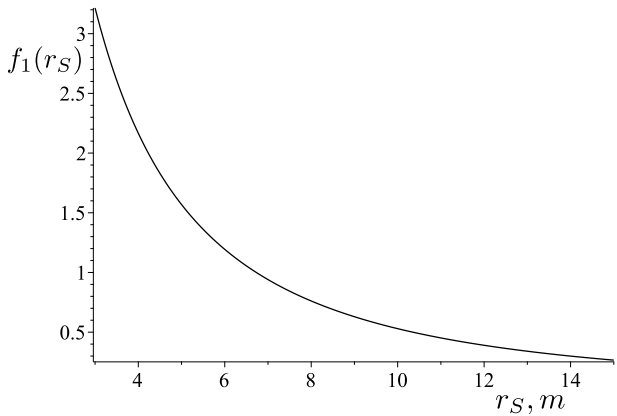}
\end{center}
\caption{Graph of the function $f_1(r_S)$ given by eq.(\ref{f1-rS}).}
\label{fig:graph2}
\end{figure}

Graph of $f_1(r_S)$ is plotted on Fig.\ref{fig:graph2}. We see that a thin luminous ring of a given radius $r_S$ and thickness $\Delta r_S$  will lead to bigger thickness of the higher-order ring in the case of smaller $r_S$.

\section{Analytical estimation of fluxes}
\label{sec: flux}

In this paper, we are dealing with a situation where the light source is in a strong gravitational field, and only the observer is in an asymptotically flat region. Due to the gravitational redshift, the frequency of the emitted photons does not equal to the frequency of the photons caught by the observer. As a result, the surface brightness of the source and the surface brightness of the image are different, see also \cite{Cunningham-1972, Cunningham-1973, Bozza-Scarpetta-2007}. We will consider the disk with radial distribution of brightness $I_S(r_S)$ and will take into account the disk rotation.

We assume that the disk is rotating, and the matter is moving in circular orbits. Then, an emitting particle at radius $r_S$ has the angular velocity $\Omega = (GM/r^3)^{1/2}$ \cite{Hobson, Luminet-1979}. For the rotating disk viewed face-on, the gravitational redshift and Doppler shift together lead to the following ratio of the frequencies \cite{Hobson, Luminet-1979}:
\begin{equation}
\frac{\omega_0}{\omega(r_S)} = \left( 1 - \frac{3m}{r_S} \right)^{1/2} \, .
\end{equation}
Here $\omega_0$ is the photon frequency measured by an observer in an asymptotically flat region, and $\omega(r_S)$ is
a photon frequency emitted by the disk near the black hole at position $r_S$.

Surface brightness $I$ of image measured by observer in the asymptotically flat region is related to the surface brightness $I_S$ of emitting source at $r_S$ as \cite{Cunningham-1972, Cunningham-1973}:
\begin{equation}
\frac{I}{I_S} = \left( \frac{\omega_0}{\omega(r_S)}      \right) ^4 = \left( 1 -      \frac{3m}{r_S}    \right)^2 \, . 
\end{equation}
Therefore, for the flux $F_n$ from $n$-th image with surface brightness $I_n$ we write:
\begin{equation} \label{F-n-01}
F_n = \int \limits_{\Delta \Omega_n} I_n(r_S) \, d\Omega   =  \int \limits_{\Delta \Omega_n} \left( 1 -   \frac{3m}{r_S}    \right)^2   I_S(r_S) \, d\Omega  \, .  
\end{equation}
To perform this integration, the solid angle $d\Omega$ can be written as $d\Omega = 2 \pi \theta d\theta$, and connection between the location $r_S$ of emitting  point of the disk and its observed position $\theta$ within $n$-th image should be taken into account.

Flux ratio of two successive rings equals to:
\begin{equation} \label{flux-ratio}
\frac{F_{n+1}}{F_n} = \frac{ \int \limits_{\Delta \Omega_{n+1}}  \left( 1 -   3m/r_S \right)^2 I_S(r_S) \, d\Omega }
{
\int \limits_{\Delta \Omega_n} \left( 1 -   3m/r_S \right)^2 I_S(r_S) \, d\Omega
}  \, .
\end{equation}
For higher-order images ($n \ge 2$), this ratio can be found analytically. From (\ref{theta-general}) and (\ref{Delta-phi}), we get that emitting point at location $r_S$  produces the point of the $n$-th higher-order image at the observed angular position $\theta_n$:
\begin{equation} \label{theta_n}
\theta_n = \frac{3\sqrt{3}m}{D_d} \left[ 1 + f(r_S) \,  e^{- \pi \left( n+ \frac{1}{2} \right)}  \right] \, ,
\end{equation}
\[
f(r_S) \,  e^{- \pi \left( n+ \frac{1}{2} \right)} \ll 1 \, .
\]
Here the angular variable $\theta_n$ changes from inner boundary $\theta^{\mathrm{in}}_n$ of $n$-th ring image to its outer boundary $\theta^{\mathrm{out}}_n$. Correspondingly, the variable $r_S$ changes from inner radius $r^{\mathrm{in}}_S$ of emitting disk to its outer radius $r^{\mathrm{out}}_S$.  

For integration over the solid angle of $n$-th ring, we can write:
\begin{equation} 
d\Omega = 2 \pi \theta_n d\theta_n \simeq  2 \pi \frac{27m^2}{D_d^2} f'(r_S) \,  e^{- \pi \left( n+ \frac{1}{2} \right)} dr_S \, ,
\end{equation}
After this transformation, we have the integration over the source surface with variable $r_S$ instead of integration over the solid angles of images with variable $\theta_n$. As a result, the integral expressions for the fluxes of rings in eq.(\ref{flux-ratio}) will differ only by the number $n$ used in the exponential factor $\exp[-\pi(n+1/2)]$ that can be taken out of the integral sign. Therefore we obtain:
\begin{equation}
\frac{F_{n+1}}{F_n} = e^{-\pi} \, .
\end{equation}
We conclude that independently of the source brightness profile $I_S(r_S)$, the ratio of higher-order ring fluxes will be equal to the ratio of their solid angles.\\

Another way to estimate fluxes can be carried out through the use of the solid angle of unlensed disk $\Delta \Omega_S$. This value is not directly observable but the estimations still can be useful because the value of $\Delta \Omega_S$ is close to the observed angular size of direct image $\Delta \Omega_0$. Let us now assume that $I_S(r_S) = \mbox{const}$.

Since $r_S \ge r_S^{\mathrm{in}}$, then
\begin{equation}
1 -  \frac{3m}{r_S}  \ge 1 -   \frac{3m}{r_S^{\mathrm{in}}}   \, .  
\end{equation}
We write:
\begin{equation} \label{F-n-02}
F_n \ge I_S \left( 1 -  \frac{3m}{r_S^{\mathrm{in}}}   \right)^2   \int \limits_{\Delta \Omega_n} d\Omega   =  I_S \left( 1 -  \frac{3m}{r_S^{\mathrm{in}}}   \right)^2  \Delta \Omega_n \, .
\end{equation}
Let us introduce the value:
\begin{equation} \label{F-S}
F_S = \int \limits_{\Delta \Omega_S} I_S \, d\Omega = I_S \Delta \Omega_S \, ,
\end{equation}
and define the magnification $\mu_n$ of $n$-th image as:
\begin{equation} \label{mu-n-def}
\mu_n = \frac{F_n}{F_S}  \, .
\end{equation}
Substituting (\ref{F-n-02}) and (\ref{F-S}) into (\ref{mu-n-def}), we find:
\begin{equation} \label{lower-limit}
\mu_n \ge \left( 1 -   \frac{3m}{r_S^{\mathrm{in}}}  \right)^2 \frac{\Delta \Omega_n}{\Delta \Omega_S} \, .
\end{equation}

Inequality (\ref{lower-limit}) represents the lower limit for $\mu_n$. Analogously, we can calculate the upper limit of $\mu_n$ using that $r_S \le r_S^{\mathrm{out}}$. Finally, we find the following double inequality which can serves as an estimation of flux magnification:
\begin{equation} \label{flux-main-result}
\left( 1 -     \frac{3m}{r_S^{\mathrm{in}}}    \right)^2  \frac{\Delta \Omega_n}{\Delta \Omega_S} \le  \mu_n \le  \left( 1 -      \frac{3m}{r_S^{\mathrm{out}}}      \right)^2  \frac{\Delta \Omega_n}{\Delta \Omega_S}  \, .
\end{equation}
To use this inequality for the estimation of flux magnification, one need to know the ratio of solid angles of corresponding $n$-image and unlensed disk.

For rings with $n \ge 2$, the ratio $\Delta \Omega_n / \Delta \Omega_S$ can be written down fully analytically. Indeed, for the solid angle of unlensed accretion disc we have
\begin{equation} \label{Omega-S}
\Delta \Omega_S = \frac{\pi}{D_d^2}  \left[ (r^{\mathrm{out}}_S)^2 - (r^{\mathrm{in}}_n)^2  \right]  \, .
\end{equation}
Combining it with eq.(\ref{Omega-n}) we find:
\begin{equation} \label{magn-geom}
\frac{\Delta \Omega_n}{\Delta \Omega_S} = 54 \, m^2 \, \frac{  f(r_S^{\mathrm{out}}) -  f(r_S^{\mathrm{in}}) }{ (r^{\mathrm{out}}_S)^2 - (r^{\mathrm{in}}_S)^2   }  \,  e^{- \pi \left( n + \frac{1}{2} \right)} \, , \; n \ge 2 .
\end{equation}\\

\section{Example of calculation}
\label{sec: example}

In this Section we present the example of calculation of angular sizes and flux magnifications of lensed images of the accretion disk. We consider the disk with inner radius $r^{\mathrm{in}}_S = 6m$ and outer radius $r^{\mathrm{out}}_S = 15m$.

Solid angle of unlensed acrretion disk can be calculated analytically by (\ref{Omega-S}):
\begin{equation}
\Delta \Omega_S \simeq 594 \, \frac{m^2}{D_d^2} \, .
\end{equation}

By substitution $n=2$ into (\ref{Omega-n}) we find analytically the solid angle of the $n=2$ higher-order ring:
\begin{equation}
\Delta \Omega_2 \simeq 0.333 \, \frac{m^2}{D_d^2} \, .
\end{equation}
We can compare the size of $n=2$ higher-order ring with size of unlensed disk by (\ref{magn-geom}):
\begin{equation} \label{Omega-2-example-analyt}
\Delta \Omega_2 \simeq 0.000560  \, \Delta \Omega_S \, .
\end{equation}

For all following higher-order rings we have the exponential decrease of the corresponding solid angles:
\begin{equation}
\Delta \Omega_{n+1} = e^{-\pi}  \Delta \Omega_n \simeq 0.0432  \, \Delta \Omega_n \;\; \mbox{for} \;\; n\ge2.
\end{equation}

For completeness, we consider also the solid angles occupied by the primary $\Delta \Omega_0$ and secondary $\Delta \Omega_1$ images. We have found these values by numerically integrating the ray trajectories. We have:
\begin{equation} \label{Omega-0-example}
\Delta \Omega_0 \simeq 1.10 \, \Delta \Omega_S \, ,
\end{equation}
\begin{equation} \label{Omega-1-example}
\Delta \Omega_1 \simeq 0.017  \, \Delta \Omega_0 \, , \quad \mbox{or} \quad \Delta \Omega_1 \simeq 0.019  \, \Delta \Omega_S \, , 
\end{equation}
\begin{equation} \label{Omega-2-example}
\Delta \Omega_2 \simeq 0.030  \, \Delta \Omega_1 \, , \quad \mbox{or} \quad  \Delta \Omega_2 \simeq 0.00051  \, \Delta \Omega_0 \, .
\end{equation}

From this example, we see that each next image is a few percent of the previous one, for all higher-order images this ratio equals to $e^{-\pi}\simeq 0.043$. This conclusion agrees with numerical and analytical results of Johnson et al \cite{Johnson-2020}. For rotating black hole, the numbers will be different.

\begin{figure}
\begin{center}
\includegraphics[width=0.48\textwidth]{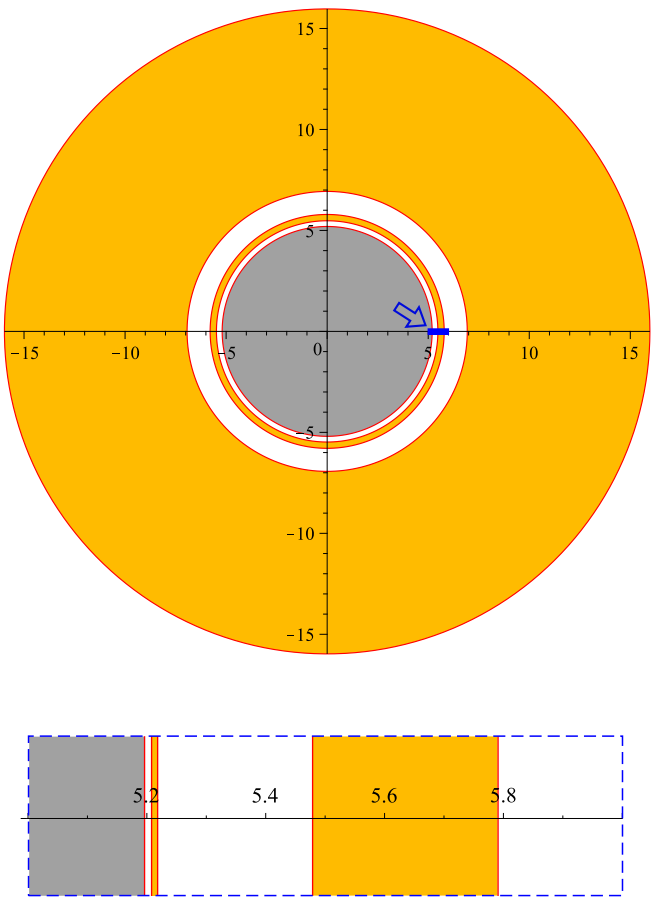}
\end{center}
\caption{Appearance of thin accretion disk around Schwarzschild black hole viewed face-on: primary image ($n=0$), secondary image ($n=1$) and $n=2$ higher-order ring. Top panel: Primary and secondary images found numerically. The disk has given inner and outer boundaries at $6m$ and $15m$ correspondingly. Inner radius equals to ISCO position, the choice of the outer radius was determined by the goals of better visualization. The primary image has form of big ring which is slightly bigger than the actual (unlensed) size of disc due to gravitational bending. Secondary image has a form of thin ring close to the edge of black hole shadow (shown by gray color). Higher-order rings are too thin to be seen in this picture, their position is shown conventionally as a red solid circle on the border of the shadow. Bottom panel: Zoomed-in image of the part of the top image, highlighted with a small blue rectangle. Part of the secondary image and part of the $n=2$ ring are visible here. The higher-order ring radii are calculated analytically by (\ref{theta-in}) and (\ref{theta-out}) with $n=2$.}
\label{fig:pr-sec-6-15}
\end{figure}

On Fig.\ref{fig:pr-sec-6-15}, we plot the calculated images of accretion disk.\\

Using (\ref{flux-main-result}), we find the estimation for magnification of $n$-ring:
\begin{equation} \label{flux-example}
 \frac{1}{4}   \,\frac{\Delta \Omega_n}{\Delta \Omega_S}  \le \mu_n \le   \frac{16}{25}   \,  \frac{\Delta \Omega_n}{\Delta \Omega_S}  \, .
\end{equation}
In particular, using (\ref{Omega-0-example}) in (\ref{flux-example}), we have:
\begin{equation} \label{flux-example-01}
 0.27  \le \mu_0 \le  0.70    \, .
\end{equation}
Using (\ref{Omega-1-example}) in (\ref{flux-example}), we have:
\begin{equation}
  0.0047  \le \mu_1 \le  0.012    \, .
\end{equation}
Using (\ref{Omega-2-example-analyt}) in (\ref{flux-example}), we obtain analytically:
\begin{equation}
  0.00014  \le \mu_2 \le  0.00036    \, .
\end{equation}\\

\section{Conclusions}
\label{sec: conclusions}

(i)  Active research of higher-order images had started about twenty years ago and was mainly focused on sources far from the black hole. The main method of analytical study is using of the gravitational lensing formalism in the strong deflection limit. Here we show that the same techniques can be applied to novel studies of higher-order ring images of luminous accretion matter around the black hole.

(ii) We consider the lensing of thin accretion disk with known inner and outer radii by the Schwarzschild black hole. The observer far from the black hole on the axis of symmetry will see the primary image of the disk (the rays go directly to the observer, number of half-orbit $n=0$), the secondary image in the form of a thin ring (the rays pass along the other side of the black hole and form an image of the back of the disk, $n=1$) and exponentially faint higher-order rings (one full revolution or more, $n\ge2$), see Fig.\ref{fig:rings}.

(iii) Our goal was to derive a completely analytical solution in closed form, for this simplified case. Using strong deflection limit technique of gravitational lensing, we find the angular radii, angular thicknesses, and solid angles of higher-order rings ($n\ge2$) in the form of compact analytical expressions. Main results are presented by formulas for inner and outer angular radii (\ref{theta-in}) and (\ref{theta-out}), angular thickness (\ref{theta-n}) and solid angle (\ref{Omega-n}) of higher-order images. Our results not only reproduce the property of exponential decrease of higher-order images, but allow one to find analytically the size of each of the rings.

(iv) The simple form of the solution allowed us to analyze it effectively and find features that might not be visible in a numerical calculation or in more detailed analytical studies that use a richer accounting of parameters. We find that the size of the higher-order rings is mainly determined by the position of the inner boundary of the accretion disk, which makes it possible to use them to distinguish between different accretion models, see Section \ref{sec: properties}.

(v) Figure with first three rings ($n=0,1,2$) together is presented, see Fig.\ref{fig:pr-sec-6-15}.

(vi) Our analytical approach provide simple calculation and comparison of the solid angles occupied by higher-order rings. Calculating of fluxes is more complicated; however, our model allows to estimate the fluxes analytically, see Section \ref{sec: flux}. Rotation of the disk is taken into account. Interestingly, the flux magnification of the primary image compared to the non-lensed disk is less than unity, see eq.(\ref{flux-example-01}), although the angular size of the primary image is slightly larger than the angular size of the unlensed disk, see eq.(\ref{Omega-0-example}). This differs from the well-known statement in gravitational lensing of distant sources, where the magnification factor for the primary image is always greater than unity. The difference is due to the fact that in case of accretion disk, the emitting sources are located near the black hole, and it becomes necessary to take into account the change in frequency and change in brightness.

(vii) Our method makes it easier to discriminate between $n=2$ and $n=3$ higher-order rings, the possible observation of which in future projects is currently being discussed.

(viii) A possible generalization of our results is the calculation of higher-order rings for the Reissner-N\"{o}rdstrom black hole, which may allow one to constraint the charge of the black hole.

(ix) In our work \cite{Tsupko-Fan-BK-2020}, we proposed using the shadow of black holes at cosmological distances as a standard ruler in cosmology. The higher-order rings are exponentially close to the edge of the shadow. Therefore, if the physical conditions around a black hole make it possible to see higher-order rings ('photon ring'), then the method of standard ruler can be implemented on the basis of observing these rings.

\section*{Acknowledgements}

The work of G.S.B.-K. and O.Yu.T. was partially supported by the Russian Foundation for Basic Research Grant No. 20-02-00455. O.Yu.T. also thanks Hanse-Wissenschaftskolleg (Institute for Advanced Study, Delmenhorst) for supporting his research stay there.

\bibliographystyle{ieeetr}

\end{document}